# The Human Blockage Impact on ARIS Assisted D2D Communication Systems

Ahmed M. Nor
*Telecommunications Department.*
*University Politehnica of Bucharest*
Bucharest, Romania
*Electrical Engineering Department*
*Aswan University*
Aswan, Egypt
ahmed.nor@upb.ro

Octavian Fratu
*Telecommunications Department.*
*University Politehnica of Bucharest*
Bucharest, Romania
octavian.fratu@upb.

Simona Halunga
*Telecommunications Department.*
*University Politehnica of Bucharest*
Bucharest, Romania
simona.halunga@upb.ro

*Abstract*—Aerial reconfigurable intelligent surface (ARIS), is an intelligent reflecting surface (IRS) mounted by unmanned aerial vehicle (UAV), represent a promising candidate for assisting device to device (D2D) millimeter wave (mmWave) communication in temporal and urgent situations, e.g., open-air events. IRS can efficiently mitigate the high blockage impact on mmWave propagation signal in base station to device use case. But, the scenario of D2D communication is different as both the transmitter (TX) to ARIS and the ARIS to receiver (RX) links are highly susceptible to be blocked due to the low height of the TX and RX. Consequently, in this paper, the impact of human bodies blockage on ARIS aided D2D mmWave communication is studied. Firstly, we assure the effectiveness of using ARIS in this network to significantly enhance its performance, then, the effect of ARIS height on the blockage occurrence and system performance is investigated to find out the optimum height. Our results proves that ARIS highly mitigates the blockage, reduces it by 85%, comparable to the case without it. Moreover, a high increase in system spectral efficiency, 1.2 bps/Hz, can be guaranteed, if ARIS is configured at optimum height.

## I. INTRODUCTION

6G networks come with many new use cases and scenarios [1]. One of them is to provide reliable and robust line of sight (LOS) links between user equipment (UEs) in device to device (D2D) communication in temporal and urgent situations, e.g., special events and disasters cases [2]. Moreover, several technologies appeared, such as intelligent reflecting surface (IRS), also called as reconfigurable intelligent surface (RIS), and unmanned aerial vehicle (UAV). Programmable RIS is a surface consisting of number of passive elements, which can be configured by changing their amplitudes and phase shifts [3], [4]. Hence, RIS can reflect the incident signal from the transmitter (TX) into the receiver (RX) direction, thus, enhancing the propagation channel, widen the network coverage and overcoming the line of sight (LOS) link blockage occurrence. On the other hand, UAV based networks have many features such as high flexibility, ease of on-demand implementation to serve in temporal situations, e.g., open-air events and urgent disasters, cost-effective for insufficient coverage areas, e.g., remote areas [5], [6].

Recently, new aerial RIS (ARIS) has been developed by mounting RIS on UAV to offer and enhance many promising characteristics comparable to fixed RIS [5], [6], e.g., wider coverage, flexible deployment, and extending reflection. Moreover, ARIS can decline the channel complexity and mitigate the interference effect in wireless communication.

Furthermore, using this solution, UAV can stop propagating signal, and just used to reflect signal by its mounted RIS, hence the energy consumption will be highly decreased. It is known that the beyond 5G bands, i.e., millimeter wave (mmWave) and terahertz (THz), highly suffer from blockage due to different obstacles. Hence, mobile RIS can be more favorable than its fixed counterpart to serve in mmWave communication to overcome blockage. Because RIS position can be optimized as well as its elements' amplitudes and phase shifts to guarantee more reliable and powerful indirect LOS links [7]. Several previous works discussed the idea of aerial RIS based networks in terms of deployment optimization [7], passive beamforming optimization [8], network performance improvement [6], channel modeling [9] and UAV detection and positioning issues [10]. Moreover, the impact of blockage between RIS and RX in RIS based system has been studied in some works [11], [12]. But, in RIS based networks, the link between TX and RIS is robust and immune against blockage, as TX and RIS are usually deployed in high positions that are not susceptible to blockage. Unlike this case, in ARIS aided device to device (D2D) communication, both TX-ARIS and ARIS-RX links can be blocked because the devices' plane is low, in range of 1m.

Consequently, in this paper, we study the effect of the human bodies blockage on ARIS assisted D2D mmWave communication system. In which, we will consider the real case, where both the TX-ARIS and ARIS-RX links can be blocked, assuming the practical Geometrical-Empirical blockage model. First, we investigate on the superior added value, because of using ARIS to assist D2D connections, comparable to networks without aerial RIS in terms of mitigating the blockage impact and enhancing the system spectral efficiency. Furthermore, we study the influence of ARIS height on the blockage occurrence and the system performance aiming to find out the optimum height within the studied use case. The remaining of the paper is organized as: Section II presents the system model and the used blockage model, meanwhile the performance evaluation is illustrated in Section III. Finally, Section IV concludes the paper.

## II. SYSTEM MODEL

Figure 1. shows the network architecture and the studied scenario of ARIS aided device to device (D2D) communication system, where a source (S) and destination (D) devices connect through the aid of the A-RIS. Here, RIS is mounted on a UAV, which can move easily to cover a wider area, e.g., stadium or public event. Both the source and destination devices have a single antenna. In this scenario, the three connection links can be blocked, first, the direct line of sight (LOS) S-D link, secondly, the S-ARIS link, and the

ARIS-D link, which are represented by red, blue, and dashed blue lines, respectively.

The configuration of ARIS is presented in Figure 2., where it is centered in the origin of x-y-z coordinates system and consists of $N$ passive reflecting elements. The amplitudes and phases of RIS elements can be configured to optimally reflect the source signal to the destination device in what is called passive beamforming (PBF), which can be done using various algorithms such as [3]. In this work, we are just interested in the performance of ARIS assisted D2D communication, hence we will consider the link model illustrated in [13], and the nonexistence of the S-D link due to blockage. The distance between the S and the ARIS center is $d_{S-ARIS}$, while the angles of elevation and azimuth of the incident beam seen by ARIS are $\theta_{S-RIS}$ and $\varphi_{S-ARIS}$, respectively. Also, the distance between the D and the ARIS center is $d_{ARIS-D}$, while the angles of elevation and azimuth of the reflected beam seen by ARIS are $\theta_{ARIS-D}$ and $\varphi_{ARIS-D}$, respectively.

This ARIS is considered as a one regime after configuring its elements amplitudes and phase shifts. Hence, the received power $P_r$ at the destination, can be presented as [13]:

$$P_r = A_r S_r \quad (1)$$

where $A_r$ refers to the receiver effective aperture and written as

$$A_r = \frac{G_r \lambda^2}{4\pi}, \quad (2)$$

where $G_r$ is the antenna gain of the D device, meanwhile $\lambda$ indicates to the free space wavelength. Also, $S_r$ is the power density at the destination which can be written as [13]

$$S_r = \frac{\frac{2P_t}{\lambda Z_R}|R|^2}{\sqrt{\left(1 + \frac{d_{ARIS-D}^2}{Z_R^2}\right)\left(1 + \frac{d_{ARIS-D}^2}{Z_R^2 \cos^4 \theta_{ARIS-D}}\right)}}, \quad (3)$$

where $P_t$ and $G_t$ are the source transmitted power and antenna gain, respectively. Here, $|R|$ refers to the common reflection amplitude of ARIS elements, while $Z_R$ is the Rayleigh length and defined as

$$Z_R = \frac{4k_o d_{S-ARIS}^2}{G_t}, \quad (4)$$

where $k_o$ is the free space wavenumber.

For the human blockage model, the one described in [14] is considered as presented in Figure 3. This Geometrical-Empirical model is a modification to Double Knife-Edge Diffraction (DKED) model to present the effect of the human body shadowing during the blocking time period. In Figure 3., the height, the length and the width of the human body are $h$, $l$ and $w$, respectively. Blockers can be in between S and D devices, S device and ARIS, and ARIS and D device, as we explained. Here, the blockers are uniformly distributed in the studied area with density $B$, which indicated by blockers per m$^2$ ($bl/m^2$). In this study, we assume all human bodies are stationary, hence, the blockage happens along all data transmission period causing a high attenuation.

III. PERFORMANCE EVALUATION

We present in this section the impact of human bodies on the blockage occurrence and the system performance in ARIS assisted wireless communication system. First, the probability of blockage and system spectral efficiency are discussed with and without the assistance of ARIS to clarify the valuable effect of using ARIS in crowded environments. Then, we study the impact of the ARIS height on the blockage and the system rate aiming to find out the optimal height for ARIS in the network. Although this study is a scenario based one, it gives an insight on the general expected performance. An urban environment, e.g., open-air events venue, with area of interest 50×50 m$^2$ is considered. Source and destination devices are randomly distributed in this area, where their transmitter and receiver work in 60 GHz mmWave band. Also, we assume the ARIS is at the center of the area, and moves vertically, though it can fly in all directions. 100,000 Monte Carlo trials are conducted for each simulation point, and the results are obtained by averaging on them. Moreover, in ARIS based network results, we assume the full blockage of S-D link. Table I. summaries other parameters.

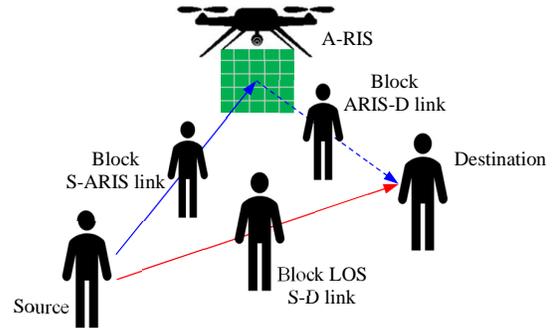

Figure 1. ARIS aided D2D mmWave network architecture.

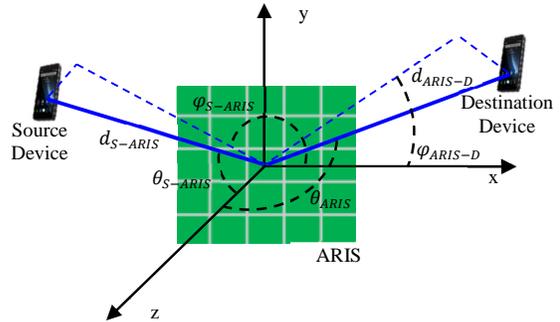

Figure 2. The configuration of ARIS link model.

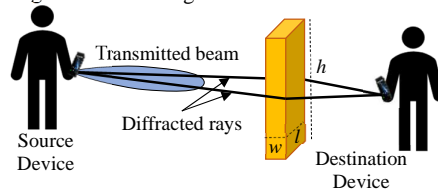

Figure 3. Human blockage model in D2D communication scenario.

TABLE I.      TABLE I. SIMULATION PARAMETERS

| Parameter | Value |
|---|---|
| Source device transmitted power, $P_t$ | 1 watt |
| Source device antenna gain, $G_t$ | 45 dB |
| Destination device antenna gain, $G_r$ | 20 dB |
| S and D heights, $h$ | **1** m |
| Human blockage dimension | **1.75**m×**0.5**m×**0.2**m |
| Operating frequency | 60 GHz |
| Common reflection amplitude, $|R|$ | 0.9 |
| Noise power | -70 dBm |

Figure 4. a, and b show the blockage occurrence probability and the spectral efficiency (SE) of the system in bps/Hz, respectively, whether the ARIS is used in the network or not, versus different blocker densities. These two figures prove how implementing ARIS can add a high value to the system, though the link between the two communication devices through the ARIS will be attenuated due to path loss. Here, we assume the vertical distance between ARIS and devices' plane is 13 m. First, ARIS aided system highly mitigates the blockage effect, e.g., with 0.2 bl/m$^2$ density, the blockage occurrence is 2% due to ARIS usage comparable to 26% without ARIS. Furthermore, with a blocker density equals 1 bl/m$^2$, an 85% reduction in blockage can be obtained using ARIS. As a result, the system with ARIS guarantees a higher spectral efficiency comparable to the system without ARIS even if the blocker density increased. Because, with higher density, the S-D link can be blocked with more than two blockers at the same time causing extremely high signal attenuation and decreasing SE. For example, ARIS aided system obtains 21.83 and 21.51 bps/Hz, if blocker densities are 0.5 and 1 bl/m$^2$, respectively, which are high increase, 3.6 and 7.4 bps/Hz, comparable to the case of without ARIS. Moreover, in denser scenario, 2 bl/m$^2$, a 124% increase in SE can be achieved.

Figure 5. presents the impact of the ARIS height on the blockage probability with three different blocker densities. The aim of this study is to give an insight on the optimum altitude of the ARIS to be able to efficiently eliminate the human bodies effect. In Figure 5., the raising of ARIS exponentially declines the blockage probability because the elevation angles between ARIS and both the S and D devices will be decreased, hence S-ARIS and ARIS-D links will have a lower susceptibility to be blocked. Moreover, it is noted that further increase in ARIS height will not remarkably affect the blockage, where this height is depends on the blocker density, e.g., it is 35 m height, if the blocker density is 1 bl/m$^2$. Consequently, we separately studied the impact of ARIS height on the system spectral efficiency in Figure 6 to define the optimum ARIS altitude. Increasing the ARIS height rises the spectral efficiency, then, after certain height it begins to decrease due to the path loss because of the increase in distance between ARIS and both S and D devices. This height is considered the optimum ARIS altitude for best system performance. For instance, with blocker densities 0.5, 1 and 1.5 bl/m$^2$, the optimum ARIS heights are 12, 14, 16 m, respectively. These optimum heights guarantee system SE of 21.86, 21.51, and 21.21 bps/Hz, respectively. Moreover, in 1 bl/m$^2$, the optimum height achieves a higher SE with 1.2 bps/Hz comparable to the height of 5 m.

## IV. CONCLUSION

In this paper, the impact of human bodies blockage on the performance of ARIS assisted D2D mmWave communication system is discussed. Here, we consider the practical Geometrical-Empirical human bodies blockage model and tame it to our scenario. First, the superior significance of implementing ARIS in the D2D communication is proven by studying the impact of blocker density on the blockage occurrence and system SE with and without ARIS. Using ARIS can guarantee a 52 and 124% increase in system SE comparable to the case without ARIS, when the blocker densities are 1 bl/m$^2$ and 2 bl/m$^2$, respectively. Furthermore, aiming to find out the optimum ARIS height, we study the blockage probability and system SE versus different ARIS height. The increase of ARIS height declines the blockage at the expense of increasing the path loss, hence the system SE is needed to be able to judge on best height. As in results, the SE increases with raising the ARIS up until reaching a certain height which is the optimum one. This height depends on the blockers' density. For instance, it is 13m vertical distance between ARIS and devices' plane, which obtains 21.5 bps/Hz SE, in case of the blocker density is 1 bl/m$^2$. Considering mobile blockers and devices in the study with a realistic mobility model can be a contribution in the future.

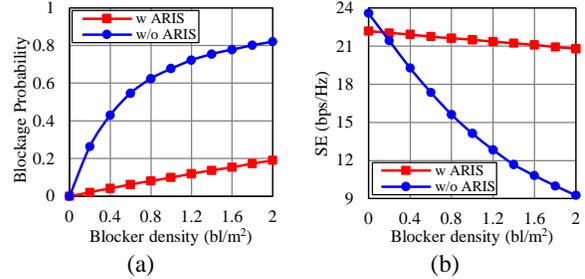

Figure 4. (a) The blockage occurrence probability, (b) the spectral efficiency, of the system with (w) and without (w/o) using ARIS.

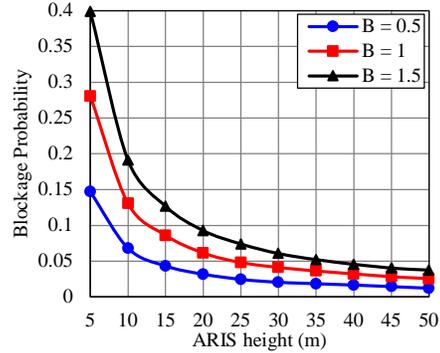

Figure 5. Blockage occurrence probability versus ARIS height with different blocker densities.

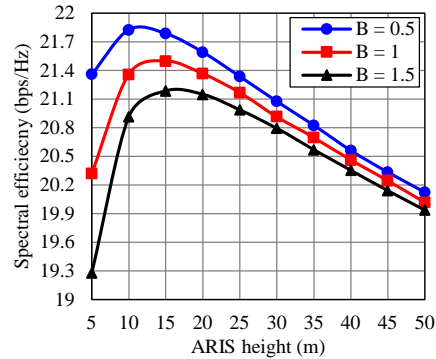

Figure 6. System spectral efficiency versus ARIS height with different blocker densities.

ACKNOWLEDGMENT

This study has been conducted under the project 'MObility and Training fOR beyond 5G eco-systems (MOTOR5G)'. This project has funded from the European Union's Horizon 2020 programme under the Marie Skłodowska Curie Actions (MSCA) Innovative Training Network (ITN) under grant agreement No. 861219.